\documentclass[entropy,preprint,article,submit,moreauthors,pdftex,10pt,a4paper]{mdpi} 

\preto{\abstractkeywords}{\nolinenumbers}

\usepackage{xspace}
\usepackage{amsmath}
\usepackage{mathrsfs} 
\usepackage{amssymb}

\firstpage{1} 
\pubvolume{xx}
\issuenum{1}
\articlenumber{5}
\pubyear{2018}
\copyrightyear{2018}

\Title{Power, Efficiency and Fluctuations in a Quantum Point Contact as Steady-State Thermoelectric Heat Engine}

\Author{Sara Kheradsoud~$^{1}$, Nastaran Dashti~$^{2}$, Maciej Misiorny~$^{2}$, Patrick P. Potts~$^{1}$, Janine Splettstoesser~$^{2}$, and Peter Samuelsson~$^{1}$*}

\AuthorNames{Sara Kheradsoud, Nastaran Dashti, Maciej Misiorny, Patrick P. Potts, Janine Splettstoesser and Peter Samuelsson}

\address{%
$^{1}$ \quad Physics Department and NanoLund, Lund University, S-221 00 Lund, Sweden \\
$^{2}$ \quad Department of Microtechnology and Nanoscience (MC2), Chalmers University of Technology, S-412 96 G\"oteborg, Sweden}

\corres{Correspondence: samuel@teorfys.lu.se}

\abstract{The trade-off between large power output, high efficiency and small fluctuations in the operation of heat engines has recently received interest in the context of thermodynamic uncertainty relations (TURs). Here we provide a concrete illustration of this trade-off by theoretically investigating the operation of a quantum point contact (QPC) with an energy-dependent transmission function as a steady-state thermoelectric heat engine. As a starting point, we review and extend previous analysis of the power production and efficiency. Thereafter the power fluctuations and the bound jointly imposed on the power, efficiency and fluctuations by the TURs are analyzed as additional performance quantifiers. We allow for arbitrary smoothness of the transmission probability of the QPC, which exhibits a close to step-like dependence in energy, and consider both the linear and the non-linear regime of operation. 
It is found that for a broad range of parameters, the power production reaches nearly its theoretical maximum value, with efficiencies more than half of the Carnot efficiency and at the same time with rather small fluctuations. Moreover, we show that by demanding a non-zero power production, in the linear regime a stronger TUR can be formulated in terms of the thermoelectric figure of merit. Interestingly, this bound
holds also in a wide parameter regime beyond linear response for our QPC device.}    


\begin{document}

\section{Introduction}

Nanoscale thermodynamics has attracted considerable attention during the last three decades. Key motivations are the prospect of on-chip cooling and power production
as well as an enhanced thermoelectric performance arising from unique properties of nanoscale systems, such as quantum size effects and strongly energy-dependent transport properties~\cite{Cahill2003,Cahill2014,Giazotto2006,Sothmann2015,Benenti2017Jun,Heremans2013Jun,Hicks1993May,Hicks1993Jun,Kim2009Feb}. Among various nanoscale systems, quantum point contacts (QPC)~\cite{vanWees1988Feb} are arguably the most simple devices which show a thermoelectric response~\cite{vanHouten1992Mar}.  A requirement of such a response is an energy-dependent transmission probability~\cite{Buttiker1990Apr,Su2012Aug}, which breaks the electron-hole symmetry. Within non-interacting scattering theory, the transmission probability fully determines the thermoelectric response of a two-terminal device. The QPC and similar devices provide a particularly interesting thermoelectric platform as their transmission probability may approximate a step function, maximizing the power generation~\cite{Whitney2014Apr,Whitney2015Mar}. This feature is in contrast  to the case of a quantum dot, where the transmission probability may approximate a Dirac delta distribution, maximizing the efficiency of heat-to-power conversion~\cite{Mahan1996Jul,Bevilacqua2018Sep,Josefsson2018,Wu2013Sep,Humphrey2005Mar}.

 The majority of previous studies on the thermoelectric properties of QPCs focused on the linear response regime~\cite{Buttiker1990Apr,vanHouten1992Mar,Butcher1990Jun,Bevilacqua2016Dec,Proetto1991Oct,Lindelof2008Apr}. In this regime, the optimal performance of thermodynamic devices was extensively investigated, especially the efficiency at maximum power which is limited by the Curzon-Ahlborn efficiency~\cite{Curzon1998Jun,VandenBroeck2005Nov,Esposito2010Oct}. 
There are however several works considering various aspects of the thermoelectric response in the non-linear regime ~\cite{Pilgram2015Nov,Nakpathomkun2010Dec,Luo2013Oct,Cipiloglu2004Sep,Dzurak1993Oct,Bogachek1998Nov,Sanchez2004Sep,Sanchez2013Jan,Whitney2014Apr,Whitney2015Mar,Whitney2013Aug,Meair2013Jan}. This includes a Landauer-B\"{u}ttiker scattering approach to the weakly non-linear regime \cite{Sanchez2013Jan,Meair2013Jan}, detailed investigations of the relation between power and efficiency  when operating the QPC as a heat engine or refrigerator~\cite{Whitney2014Apr,Whitney2015Mar,Whitney2013Aug,Meair2013Jan} as well as the full statistics of efficiency fluctuations \cite{Pilgram2015Nov}.

Here, we review the thermoelectric effect of a QPC acting as a steady-state thermoelectric heat engine. We focus on the non-linear response regime by fully accounting for gauge invariance in the QPC transmission function. Moreover, we analyze the output power and the efficiency for different parameter regimes, varying the smoothness of the step in the transmission probability of the QPC. In addition to a high efficiency and power production, a small amount of fluctuations is desirable in the output of a heat engine. However, these three quantities, which we will analyze as three independent performance quantifiers, are often restricted by a thermodynamic uncertainty relation (TUR), preventing the design of an efficient and powerful heat engine with little fluctuations \cite{Barato2015,Gingrich2016,Pietzonka2017,Horowitz2017,Pietzonka2018,Seifert2018}. 
In this paper, we use a TUR-related coefficient as an additional combined performance quantifier, accounting for power output, efficiency, and fluctuations together.
While TURs have been rigorously proven for time-homogeneous Markov jump processes with local detailed balance \cite{Gingrich2016,Horowitz2017}, they are not necessarily fulfilled in systems well described by scattering theory \cite{Agarwalla2018Oct}. Nevertheless, we find the TUR to be valid in a temperature- and voltage-biased QPC. We note that recently, it has been shown that a weaker, generalized TUR applies whenever a fluctuation theorem holds \cite{Hasegawa2019Feb,Potts2019}. Here, we show further constraints on the TUR under the restriction that the thermoelectric element \textit{produces} power, necessary to define a useful performance quantifier. Interestingly, in linear response, this constraint can be related to the figure of merit, $ZT$. 

This paper is structured as follows. In Sec.~\ref{sec_sys}, we introduce the model of a QPC with smooth energy-dependent transmission, as well as the transport quantities and resulting performance quantifiers of interest. The latter are then analyzed for the QPC with different degrees of smoothness of the transmission function, namely, the output power in Sec.~\ref{sec_power}, the efficiency in Sec.~\ref{sec_efficiency}, the (power) fluctuations in Sec.~\ref{sec_fluc}, and the combined performance quantifier deduced from the TUR in Sec.~\ref{sec_TUR}.

\section{Model system and transport theory}\label{sec_sys}

\begin{figure}[b!!!]
\centering
\includegraphics[scale=1]{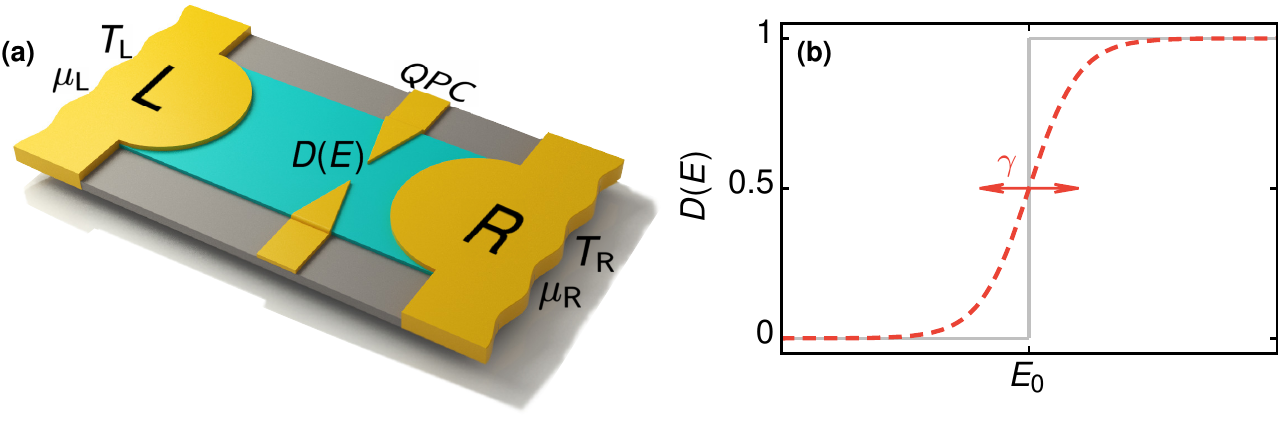}
\captionof{figure}{(a) Schematic depiction of the system, with a quantum point contact (QPC) connected to two electronic reservoirs, L and R, with chemical potentials $\mu_\text L$ and $\mu_\text{R}$ and temperatures $T_\text{L}$ and $T_\text{R}$, respectively. (b)~Transmission probability~$D(E)$ shown as a function of energy, Eq.~(\ref{eq:DE_def}), with a step positioned at energy $E_0$ and energy-smearing width, or smoothness, $\gamma$. The solid line shows the limit of vanishing width~\mbox{$\gamma\rightarrow 0$}.\label{fig:transmission}}
\end{figure}

We consider the two-terminal setup shown in Fig.~\ref{fig:transmission}, with a single-mode quantum point contact (QPC) connected to a left (L) and a right (R) electronic reservoir, characterized by  chemical potentials \mbox{$\mu_\text{L}=\mu_0-eV_\text{L}$} and \mbox{$\mu_\text{R}=\mu_0-eV_\text{R}$}, and kept at temperatures $T_\text{L}=T_0$ (cold reservoir) and $T_\text{R}=T_0+\Delta T$ (hot reservoir), respectively. Here, $V_\text L$ and $V_\text R$ are externally applied voltages, $\mu_0$ denotes the chemical potential in the absence of voltage bias, $T_0$ corresponds to the background temperature and $\Delta T\geq 0$ stands for the temperature difference due to heating of the right reservoir. In the following, we always set $\mu_0$ as the reference energy.

\subsection{Quantum point contact}
We employ the established model for a QPC \cite{Buttiker1990Apr} and describe the energy-dependent transmission probability as
\begin{equation}\label{eq:DE_def}
 D(E)=\frac{1}{1+\exp\!\Big(\frac{-E+E_0}{\gamma}\Big)}\ .
\end{equation}
This is a step-like function of the energy $E$, see Fig.~\ref{fig:transmission}(b), where $E_0$ and $\gamma$ denote the position and width in energy of the step, respectively. For a vanishingly small width, $\gamma \rightarrow 0$, the transmission probability reduces to a step function, \mbox{$D(E)\rightarrow \theta(E-E_0)$}. 

In experiments with 2DEGs, the width or smoothness of the QPC barrier $\gamma$, typically takes values of the order of 1~meV (corresponding to temperatures of the order of 10K)~\cite{vanHouten1992Mar,Dzurak1993Oct,Waldie2015Sep,Taboryski1995Jan}. The results presented in this paper are equally valid for different types of conductors, where the transmission function has a (smooth) step-like behavior, such as quantum wires with interfaces or controlled by finger gates. Here, smoothness parameters $\gamma$ of values down to several $\mu e$V are expected (corresponding to temperatures of the order of 10-100mK)~\cite{Nakpathomkun2010Dec}.

\subsection{Non-linear transport theory}
The transport properties of the system are described by scattering theory~\cite{Butcher1990Jun}. In the non-linear transport regime the scattering properties of the QPC become dependent on the applied voltage~\cite{Christen1996Sep} and possibly also on the temperature bias~\cite{Sanchez2013Jan,Sanchez2016Dec,Benenti2017Jun}. Since the details of this dependence will not be of importance for our analysis, we for simplicity consider a basic model with the QPC-potential capacitively coupled with equal strength to the two terminals L and R. This leads to a modification of the transmission probability as 
\mbox{$D(E)\rightarrow D(E+e[V_\text R+V_\text L]/2)$}, 
guaranteeing a gauge-invariant formulation of the problem with observable quantities only dependent on the potential difference \mbox{$V=V_\text L-V_\text R$}. Performing a suitable shift of the energy variable we can then write the   average currents of interest, for charge, $I_\alpha$, and heat, $J_\alpha$, that are flowing out of reservoir $\alpha$ as 
\begin{equation}
 I_\alpha=-\frac{e}{h} \int_{-\infty}^{\infty}dE\,  D(E) \big[f_\alpha (E)-f_{\overline{\alpha}} (E) \big],
 \label{chargecurr}
\end{equation}
and 
\begin{equation}
 J_\alpha=\frac{1}{h} \int_{-\infty}^{\infty}dE\, (E+\tau_{\alpha} eV/2) D(E) \big[f_\alpha(E)-f_{\overline{\alpha}}(E) \big],
 \label{heatcurr}
\end{equation}
where $\overline{\alpha}$ should be understood as follows: $\overline{\text{L}}=\text{R}$ and $\overline{\text{R}}=\text{L}$, whereas $\tau_{\text L}=1$ and $\tau_{\text R}=-1$. Here, we have introduced the modified Fermi distribution functions~$f_\alpha(E)$ already taking into account the energy-shift performed above,
\begin{equation}
f_\alpha(E)=\left[1+\exp\!\left(\frac{E+\tau_{\alpha} eV/2}{k_\text B T_\alpha}\right)\right]^{-1} 
\quad
\text{for}
\quad \alpha=\text{L,R}.
\end{equation}
While current conservation ensures $I_{\rm L}=-I_{\rm R}\equiv I$, energy conservation results in $J_{\rm L}=-J_{\rm R}-IV$.

To analyze the fluctuations in the system we also need the zero-frequency charge-current noise, given by~\cite{Blanter2000Sep}
\begin{align}
S_I=\frac{e^2}{h}\int_{-\infty}^{\infty}dE \Big\{&D(E)\left[f_\text{L}(E)\big(1-f_\text{L}(E)\big)+f_\text{R}(E)\big(1-f_\text{R}(E)\big)\right] \\ \nonumber
 +\ & D(E)\big[1-D(E)\big]\big[f_\text{L}(E)-f_\text{R}(E)\big]^2\Big\}.
\end{align}
In addition to the study of the noise, it is often convenient to analyze the Fano factor
\begin{equation}\label{eq_fano}
    F=\frac{S_I}{|2eI|},
\end{equation}
being a measure of how much the noise deviates from the one of Poissonian statistics (for which~\mbox{$F=1$}).

\subsection{Thermodynamic laws and performance quantifiers}
 The laws of thermodynamics set very general constraints on the quantities introduced above and on the performance quantifiers, which we are going to study in this paper. We start by introducing the thermodynamic laws, which can be proven to hold in scattering theory \cite{Benenti2017Jun}. 
 The first law of thermodynamics guarantees energy conservation and can be written as
\begin{equation}
    \label{eq:firstlaw}
    J_L+J_R = P.
\end{equation}
 Here, we have introduced the electrical power produced,
\begin{equation}
P=-VI,
\label{poweq}
\end{equation}
where $-VI>0$ if the current flows against the applied bias.
The second law of thermodynamics states that the entropy production $\sigma$ is non-negative. In our two-terminal geometry, it can be written as
\begin{equation}
    \label{eq:secondlaw}
    \sigma = -\frac{J_L}{T_L}-\frac{J_R}{T_R}\geq 0.
\end{equation}
This expression  determines the direction of energy flows through the system. It equals zero in case that a process is \textit{reversible}.

 To determine the performance of the QPC as a heat engine, we now consider three independent quantities and combine them with each other.
 The first performance quantifier is given by  the electrical power, Eq.~(\ref{poweq}), which following the first law, Eq. (\ref{eq:firstlaw}), is fully produced from heat.
 
The second performance quantifier we consider is given by the efficiency
\begin{equation}\label{eq_eff_def}
\eta=\frac{P}{J_\text{R}}=-\frac{VI}{J_\text{R}},
\end{equation}
where $J_\text{R}$ is the heat current that flows out of the hot reservoir. As long as power is positive, the efficiency is bounded by the second law of thermodynamics, Eq.~(\ref{eq:secondlaw}),
\begin{equation}
    \label{eq:efficiencybound}
    0\leq \eta \leq \eta_\text C
    \quad
    \text{with}
    \quad
    \eta_\text C=1-\frac{T_L}{T_R}=\frac{\Delta T}{T_0+\Delta T},
\end{equation} 
where $\eta_\text C$ denotes the Carnot efficiency. The dissipation arising from an inefficient heat to work conversion is quantified by the entropy, which thereby relates efficiency and produced power to each other
\begin{equation}
    \label{eq:entropyprod}
    \sigma 
    = 
    \frac{P}{T_0}
    \cdot
    \frac{\eta_{\rm C}-\eta}{\eta}.
\end{equation}

It is desirable to have a heat engine which not only produces large power, at high efficiency, but also  with low fluctuations. The third independent performance quantifier is therefore provided by the low-frequency power fluctuations
\begin{equation}
S_P=V^2S_I.
\label{powfluct}
\end{equation}

Interestingly, a trade-off between these quantities in the form of a TUR usually exists, as discussed in more detail in Sec.~\ref{sec_TUR}. This trade-off is typically written in the form of
\cite{Barato2015,Gingrich2016}
\begin{equation}
    \label{TUR1}
   Q_\text{TUR}
   \equiv 
   \frac{I^2}{S_I}
   \cdot
   \frac{k_{\rm B}}{\sigma}\leq\frac{1}{2},
\end{equation}
where we have introduced the coefficient $Q_\text{TUR}$. While this inequality is not always fulfilled for systems well described by scattering theory \cite{Agarwalla2018Oct}, we find it to be respected in our system for all parameter values. Importantly this coefficient can be cast into the form \cite{Pietzonka2018}
\begin{equation}
 Q_\text{TUR}
 = 
 P
 \frac{\eta}{\eta_c-\eta}
 \cdot
 \frac{k_\text BT_0}{S_P},
 \label{TUR2}
\end{equation}
where we used Eqs.~\eqref{eq:entropyprod} and \eqref{powfluct}. Thus, under the constraint of positive power production and efficiency, we identify $Q_\text{TUR}$ as a convenient combined performance quantifier, accounting for  power production, efficiency and power fluctuations together, where $1/2$ sets the optimum value.

\subsection{Linear-response regime}
In order to compare with the much more studied linear-transport regime, we here present the relevant transport properties in this limit. Specifically, with small applied voltage and thermal bias, we can write the heat and charge current on the convenient matrix form \cite{Butcher1990Jun} as
 \begin{equation}
\left(\begin{array}{c}  
        I \\
        J
       \end{array}\right)=
       \left(\begin{array}{cc}  
        G & L\\
        M & K
       \end{array}\right)
       \left(\begin{array}{c}  
        V \\
        \Delta T
       \end{array}\right),
       \label{linresp}
    \end{equation}
where (only due to linear response!) $J=J_\text{L}=-J_\text{R}$, and the matrix elements are defined as
\begin{equation}
    G=\frac{e^2}{h}\,\mathcal{I}_0, 
    \quad 
    L=-\frac{M}{T_0}=\frac{e}{h} k_\text{B}\, \mathcal{I}_1, 
    \quad 
    K=-\frac{1}{h}  (k^2_\text{B}T_0)\, \mathcal{I}_2, 
\end{equation}
with
\begin{equation} 
    \mathcal{I}_n
    = 
    \int_{-\infty}^{\infty}dE\, 
    D(E)   
    \bigg(\frac{E}{k_\text{B}T_0} \bigg)^{\!\!n} 
    \bigg(-\frac{\partial
	f_0(E)}{\partial E} \bigg). 
\end{equation}
In the same limit, the charge current noise reduces to the equilibrium noise, given by $S_I=2k_\text{B}T_0G$, in accordance with the fluctuation-dissipation relation.

Another performance quantifier, which is often used in the linear response, is the figure of merit $ZT$. It is given by
\begin{equation}
    ZT=\frac{L^2}{GK-L^2T_0}T_0,
    \label{figofmerit}
\end{equation}
in terms of the response coefficients given above.

\section{Power production}\label{sec_power}
\begin{figure}
\centering
\includegraphics[scale=1]{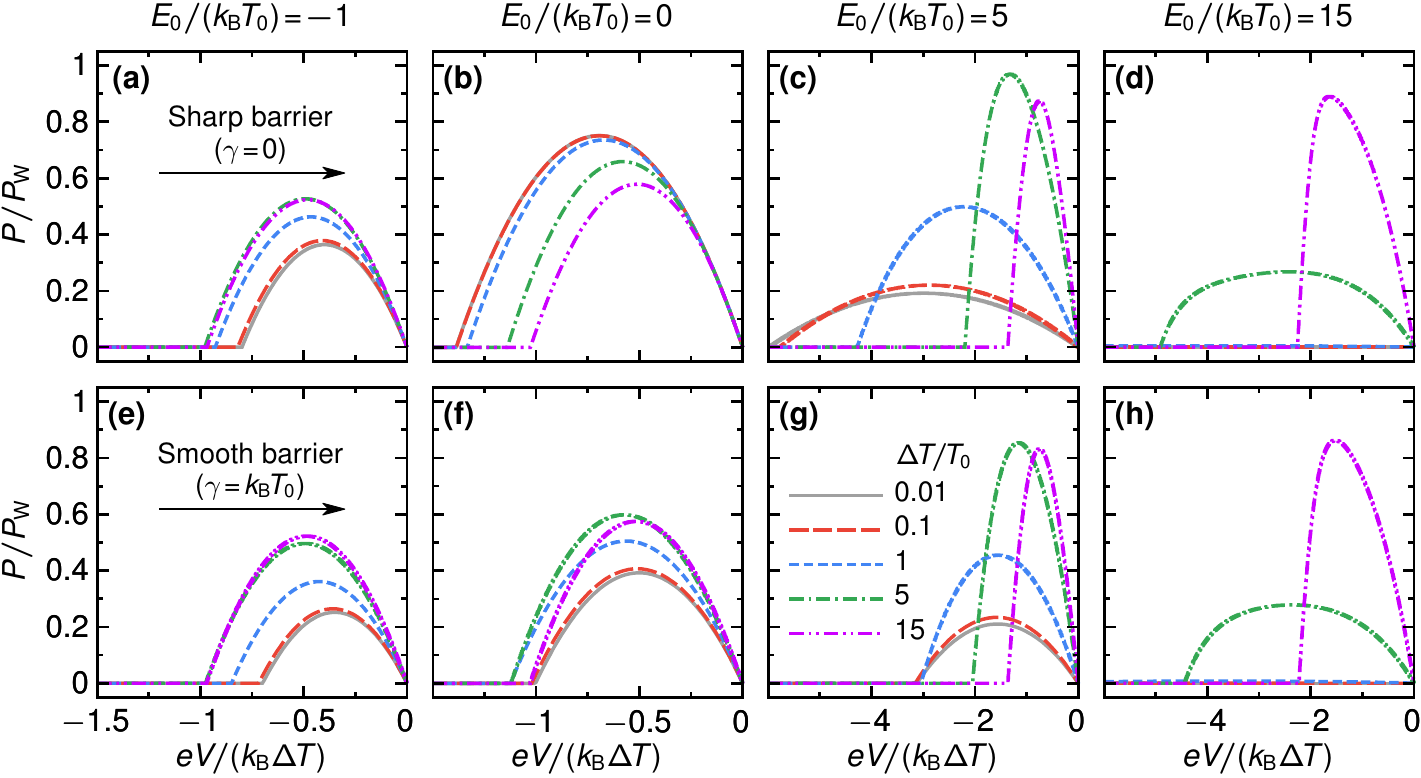}
\captionof{figure}{Power $P$, normalized by the power bound $P_\text{W}$, defined in Eq.~(\ref{eq:PWhitney}), as a function of applied bias $eV/(k_\text B \Delta T)$ for a set of step energies $E_0/(k_\text B T_0)$, shown in different columns, and thermal bias values $\Delta T/T_0$, represented by different styles of lines [the same for all panels, see inset in (g)]. 
Panels (a)-(d) [(e)-(h)] correspond to transmission functions with a step smoothness of  $\gamma\rightarrow 0$ [$\gamma=k_\text B T_0$].   
\label{fig:pow-line}}
\end{figure}

To characterize the performance of the engine, we first consider the power $P$ produced. The power as a function of applied bias $V$, for different values of the step energy $E_0$ and temperature bias $\Delta T$, is shown in Fig. \ref{fig:pow-line} for both sharp ($\gamma \rightarrow 0$) and smooth ($\gamma=k_\text B T_0$) transmission step. 
As is seen from the figure, a common feature for all P-vs-V curves is that they first increase monotonically from $P=0$ at $V=0$ with increasing negative voltage. At some voltage $V_\text{max}$ the power reaches its maximal value, $P^{\text V}_\text{max}$, and then decreases monotonically to zero, reached at the stopping voltage $V_\text s$. The maximum power with respect to voltage is a function of $E_0/(k_\text B T_0),\Delta T/T_0$ and $\gamma/(k_\text B T_0)$, i.e.,
\begin{equation}
P^{\text V}_\text{max}= P^{\text V}_\text{max}\left(\frac{E_0}{k_\text B T_0},\frac{\Delta T}{T_0},\frac{\gamma}{k_\text B T_0}\right).   
\end{equation}
In addition, we note that in the linear-response regime we have $P_\text{max}^\text V=[L^2/(4G)]\Delta T^2$ with $V_\text{max}=V_\text s/2=-(L/[2G])\Delta T$. From Fig. \ref{fig:pow-line}, it is clear that the power as a function of voltage depends strongly on all parameters $E_0/(k_\text B T_0),\Delta T/T_0$ and $\gamma/(k_\text B T_0)$. In particular, going from the linear to the non-linear regime by increasing $\Delta T/T_0$, the maximum power $P^\text V_\text{max}$ might increase or decrease depending on the step properties $\gamma$ and $E_0$.

\subsection{Maximum power}
\begin{figure}
\centering
\includegraphics[scale=1]{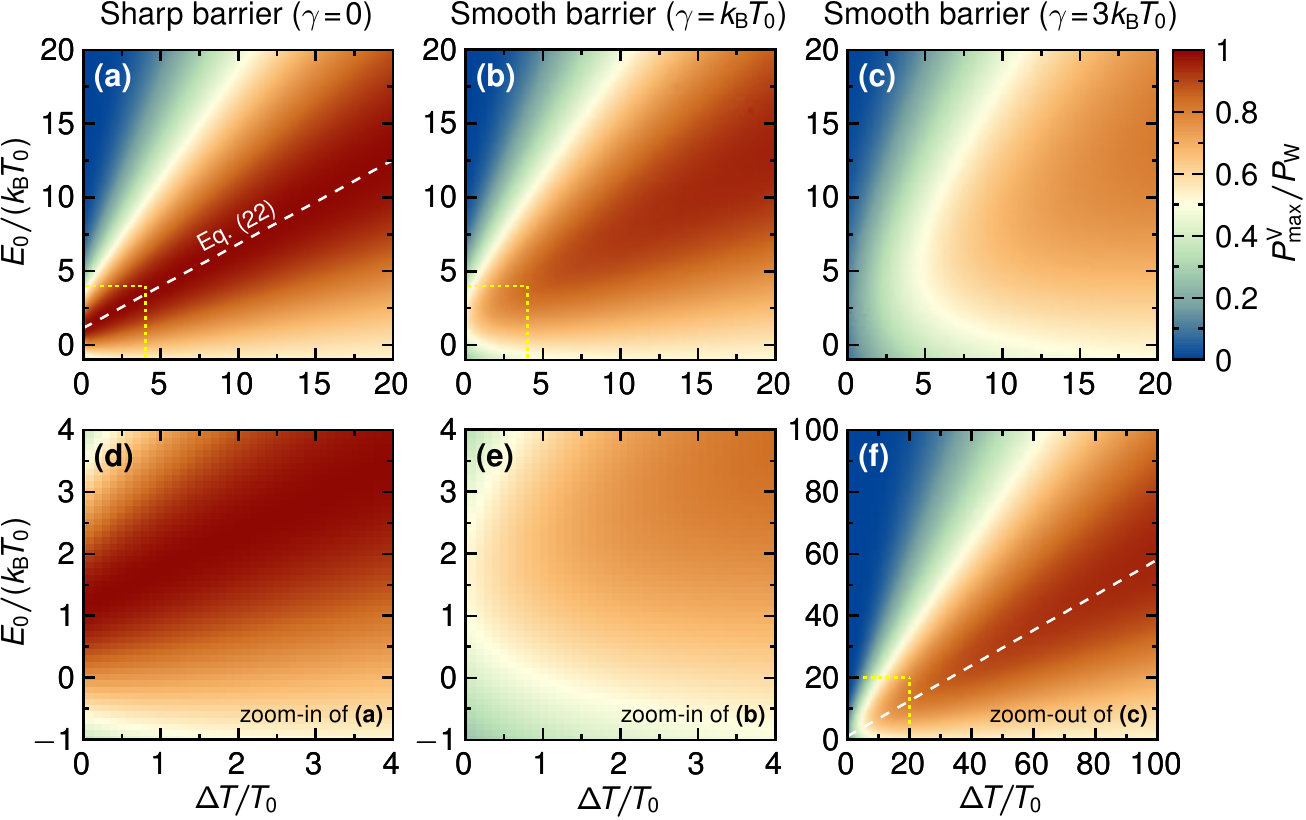}
\captionof{figure}{Maximum power with respect to voltage, $P_\text{max}^\text{V}$, as a function of  $E_0/(k_\text B T_0)$ and $\Delta T/T_0$ presented for three different values of the step smoothness $\gamma/(k_\text B T_0)=0,1,3$ [panels (a)-(c)]. 
The white dashed lines in panel (a) and (f) illustrate Eq.~(\ref{powline}). Panels (d) and (e) show close-ups of regions in panels (a) and (b), respectively, indicated with yellow dotted rectangles. On the other hand, panel (f) corresponds to an extended parameter regime of panel (c).
\label{fig:powmax2D}}
\end{figure}

To further analyse the properties of $P^{\text V}_\text{max}$, we first recall from the seminal work of Whitney \cite{Whitney2013Mar,Whitney2014Apr,Whitney2015Mar} that the power is bounded from above by quantum mechanical constraints. It was shown that the upper bound is reached for a QPC with a sharp step, $\gamma \rightarrow 0$, for which, using Eqs.~(\ref{chargecurr}) and  (\ref{poweq}), the power becomes
\begin{equation}
P_\text{sharp}
=
-\dfrac{(k_\text B T_0)^2}{h}
\cdot
\frac{eV}{k_\text B T_0}
\left\{
\frac{eV}{k_\text B T_0}-\left(1+\frac{\Delta T}{T_0}\right)\ln\left[f_{\text R}(E_0)\right]+\ln \left[f_{\text L}(E_0)\right] 
\right\}.
\label{powsharp}
\end{equation}
Maximizing this expression with respect to $eV/(k_\text B T_0)$ and $E_0/(k_\text B T_0)$ we find that the maximizing voltage is given by \mbox{$eV_\text{max}=-\xi k_\text B \Delta T$} where $\xi\approx 1.14$ is the solution of $\ln(1+e^{-\xi})=-\xi e^{-\xi}/(1+e^{-\xi})$~\cite{Benenti2017Jun}. Moreover, the  maximizing step energy $E_{0,\text{max}}$ and temperature bias $\Delta T_\text{max}$ are related via \cite{Whitney2014Apr,Pilgram2015Nov}
\begin{equation}
\frac{E_\text{0,max}}{k_\text{B}T_0}=\xi \left(1+\frac{\Delta T_\text{max} }{2T_0}\right).
\label{powline}
\end{equation}
Inserting this expression, together with the relation for the maximizing voltage, into Eq. (\ref{powsharp}) we reach the upper bound for the power established by Whitney \cite{Whitney2013Mar} and related to the Pendry bound~\cite{Pendry1983Jul},
\begin{equation} \label{eq:PWhitney}
P_\text{W}=-\frac{(k_\text B\Delta T)^2}{h}\, \xi\, \ln\left(1+e^{\xi}\right) \approx 0.32\, \frac{(k_\text B\Delta T)^2}{h},
\end{equation}
which, we emphasize, holds in the linear as well as in the non-linear regime. To relate to this upper bound, in Fig. \ref{fig:powmax2D}(a)-(c) we present a set of density plots of $P^{\text V}_\text{max}$ as a function of $E_0/(k_\text B T_0)$ and~$\Delta T/T_0$ for different values of  step smoothness parameters $\gamma$. 
%
From the figure it is clear that, for a sharp step, $\gamma \rightarrow 0$, there is a broad range of $E_0/(k_\text B T_0)$ and $\Delta T/T_0$ around the dashed line in the $(E_0,\Delta T)$-space, given by Eq. (\ref{powline}), for which $P_\text{max}^\text V$ is close to the theoretical maximum value~$ P_\text{W}$. For a step smoothness up to $\gamma \sim k_\text B T_0$, the situation changes only noticeably for small $\Delta T/T_0$. This is illustrated clearly in the close-ups in panels (d)-(e) of Fig.~\ref{fig:powmax2D} . Increasing the smoothness further, the region with maximum power close to $P_\text{W}$ shifts to higher values $E_0$ and~$\Delta T$, however, still largely centered around Eq.~(\ref{powline}), as is shown in Fig.~\ref{fig:powmax2D}(f).

To provide a more quantitative analysis of this behaviour, below we investigate two limiting cases for~$\gamma$ in further detail.

\subsubsection{Small smoothness parameter $\gamma/(k_\text B T_0) \ll 1$}
In the limit, where the value of the smoothness parameter $\gamma$ is small, $\gamma/(k_\text B T_0) \ll 1$, the expression for the transmission probability in Eq.~(\ref{eq:DE_def}) can be expanded to leading order in $\gamma$ as \cite{Lukyanov95}
\begin{equation}
D(E)=\theta(E-E_0)+\gamma^2\frac{\pi^2}{6}\cdot\frac{d}{dE}\delta(E-E_0).
\end{equation}
Inserting this into the expression for the charge current, Eq. (\ref{chargecurr}), and performing a partial integration for the delta function derivative, we get the power 
\begin{equation}
P=P_\text{sharp}- \gamma^2 \frac{eV}{h}
\cdot
\frac{\pi^2}{6}
\cdot
\frac{d}{d E_0} 
\big[f_\text{L}(E_0)-f_\text{R} (E_0) \big],
\label{powgam}
\end{equation}
with $P_\text{sharp}$ given in Eq.~(\ref{powsharp}).
To estimate how the overall maximum power is modified due to finite smoothness we insert into Eq.~(\ref{powgam}) the values for $eV/(k_\text B T_0)$, $E_0/(k_\text B T_0)$ and $\Delta T/T_0$ along the line in the $(E_0,\Delta T)$-space, see Fig. \ref{fig:powmax2D}, which gives the bounded power for the sharp barrier. We find
\begin{equation}
P(E_\text{0,max},V_\text{max})
=
P_\text{W}
\bigg\{
1- 1.06 \bigg( \frac{\: \gamma }{k_\text{B}T_0} \bigg)^{\!\!2} \frac{1}{1+\Delta T/T_0} \bigg\},
\end{equation}
noting that $\Delta T$ is related to $E_\text{0,max}$ via Eq. (\ref{powline}). This expression quantifies the effect of the barrier smoothness visible in Fig. \ref{fig:powmax2D}, namely, that the maximum power $P_\text{max}^\text V$ in the region along the line in the $(E_0,\Delta T)$ plane defined by Eq. (\ref{powline}) is mainly affected for small $\Delta T/T_0$, and approaches $P_\text{W}$ in the strongly non-linear regime, $\Delta T/T_0 \gg 1$. 

\subsubsection{Smoothness $\gamma=k_\text B T_0$}
Also in the case, where the barrier gets smoother, such that $\gamma$ equals the base temperature, $\gamma=k_\text B T_0$, we can perform an analytical investigation. Focusing on the linear response regime, $\Delta T/T_0 \ll 1$, where the effect of the smoothness is most pronounced, we can write the power in a compact form as
\begin{align}\label{eq:P_def_linresp_gT0}
P
=
-
\frac{eV}{h}
\bigg\{
\bigg[
-\mathcal{N}(E_0)
-
E_0
\,
\frac{d\mathcal{N}(E_0)}{dE_0}
\bigg]
eV
-
\frac{1}{2}
E_0^2
\,
\frac{d\mathcal{N}(E_0)}{dE_0}
\cdot
\frac{\Delta T}{T_0}
\bigg\}
.
\end{align}
where $\mathcal{N}(E)$ is the Bose-Einstein distribution function, $\mathcal{N}(E)
=\big\{\exp[E/(k_\text{B} T_0)]-1\big\}^{-1}$. As discussed above, in the linear response regime the maximizing voltage is $V_\text{max}=V_\text{s}/2$, where the stopping voltage $V_\text{s}$ is the voltage that makes the expression in the curly bracket in Eq. (\ref{eq:P_def_linresp_gT0}) vanish. Further maximizing over $E_0$ then gives $E_\text{0,max}=1.6\, k_\text{B} T_0$, which inserted into the power expression gives
\begin{equation}
P_\text{max}^{\text{V},E_0} \approx 0.5\, P_\text{W}.
\end{equation}
From Fig.~\ref{fig:powmax2D} it is clear that both $E_\text{0,max}$ and $P_\text{max}^{\text{V},E_0}$ are in good agreement with the numerical result.

\section{Efficiency}\label{sec_efficiency}
   
Taking into account the aspect of limited resources, the power output is often not the most significant performance quantifier. A more relevant quantity is then the efficiency of a device. For a heat engine, it is defined as the power output divided by the heat absorbed from the hot bath, Eq.~\eqref{eq_eff_def}.

\begin{figure}
\centering
\includegraphics[scale=1]{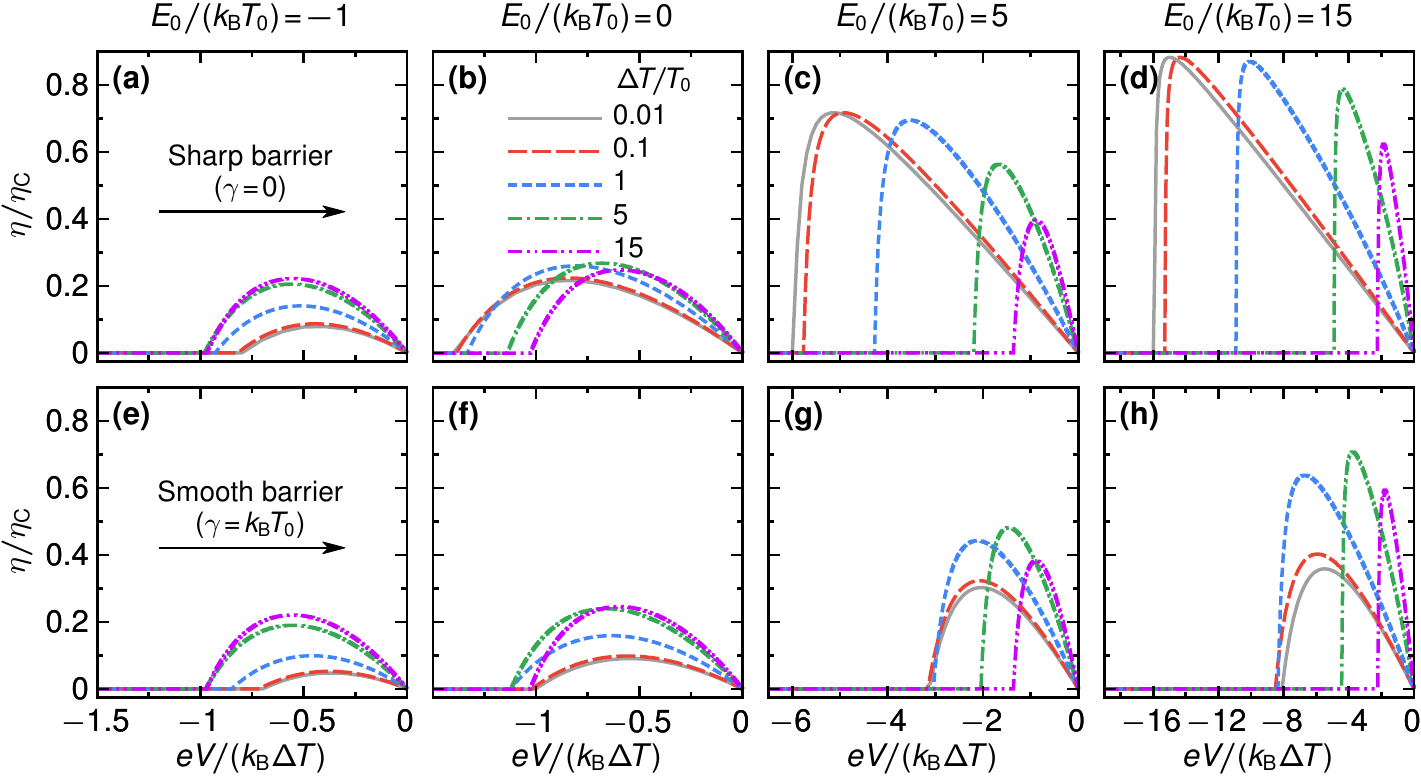}
\captionof{figure}{Efficiency as function of voltage for a sharp barrier (top row) and for a smooth barrier,  $\gamma=k_\text{B}T$ (bottom row), for selected temperature gradients $\Delta T$ (see different lines) and step energies $E_0$ (see different columns).
\label{fig:eff}}
\end{figure}

We show the efficiency of the QPC as a steady-state thermoelectric heat engine in Fig.~\ref{fig:eff}. Panels~(a) to~(d) show the efficiency for the sharp barrier as function of voltage~$eV/(k_\text{B}\Delta T)$ for different temperature gradients~$\Delta T/T_0$ and step energies $E_0/(k_\text{B}T_0)$. For small absolute values of the step energies, see panels (a) and (b) for two examples with $E_0/(k_\text{B} T_0)=-1,0$, the efficiency is rather small with respect to the Carnot efficiency, $\eta/\eta_\text{C}\lessapprox 0.25$ and its overall shape only weakly depends on the temperature gradient. This is radically different for larger values of $E_0$: panels (c) and (d) of Fig.~\ref{fig:eff} show a strong increase of the efficiency, which for $E_0/(k_\text{B}T_0)=15$ and large temperature gradients can reach about $90\%$ of the Carnot efficiency. Also the stopping voltage $V_\text{s}$, at which the efficiency is zero and the device stops working as a thermoelectric, is strongly increased, depending on the temperature gradient. 

For large $E_0$, see panel (d) of Fig.~\ref{fig:eff}, and small temperature gradients, where large maximum efficiency values are reached, the efficiency-voltage relation takes a close-to-triangular shape. In this regime, we have that $E\pm eV/2\gg T_0,T_0+\Delta T$ for all energies above the step energy $E_0$. Therefore, only the \textit{tails} of the Fermi functions contribute in Eqs.~(\ref{chargecurr}) and (\ref{heatcurr}) and the efficiency in linear response in $\Delta T$ can be approximated~as
\begin{equation}\label{eq_triangle}
 \eta=-\frac{eV}{E_0}\,\theta\!\left(eV+E_0\frac{\Delta T}{T_0}\right)\ .
\end{equation} 
This formula describes well the triangular shape of the curves in panel~(d), including  the stopping voltage at small $\Delta T$ and large $E_0$, given by $eV_\text{s}/k_\text{B}\Delta T\approx -E_0/k_\text{B}T_0$, from the argument of the Heaviside function $\theta$ in Eq. (\ref{eq_triangle}). We note that for $V \rightarrow V_\text s$ the efficiency $\eta \rightarrow \Delta T/T_0\approx\eta_{\text C}$, i.e., the efficiency approaches the Carnot limit, see Eq.~(\ref{eq:efficiencybound}). The mechanism for this is the same as described in Ref. \cite{Humphrey2005Mar}; transport effectively takes place in a very narrow energy interval around $E_0$, where the distribution functions $f_\text L(E_0)\approx f_\text R(E_0)$. 

Panels (e) to (h) of Fig.~\ref{fig:eff} show results for the changes in the efficiency for a smooth barrier, $\gamma=k_\text{B}T_0$. At temperature gradients that are much larger than the smoothness---here the case for $k_{\rm B}\Delta T/\gamma=5,15$---the results for the efficiency are very similar to the case of the sharp barrier. This is in agreement with the discussion on the power production in the previous section, Sec.~\ref{sec_power}. At small temperature gradients, however, the efficiency gets strongly reduced by the effect of the smoothness. This is particularly striking for large step energies, see panels~(g)-(h) for $E_0/k_\text{B}T_0=5,15$, respectively. Here, efficiencies that were close to Carnot efficiency for a sharp barrier get reduced by a factor three due to the barrier smoothness. The reason is that increasing smoothness leads to a broadening of the energy interval where the transport takes place, and hence a breakdown of the mechanism for Carnot efficiency discussed in Ref. \cite{Humphrey2005Mar}.

\subsection{Maximum efficiency}

\begin{figure}[t]
\begin{center}
\centering
\includegraphics[scale=1]{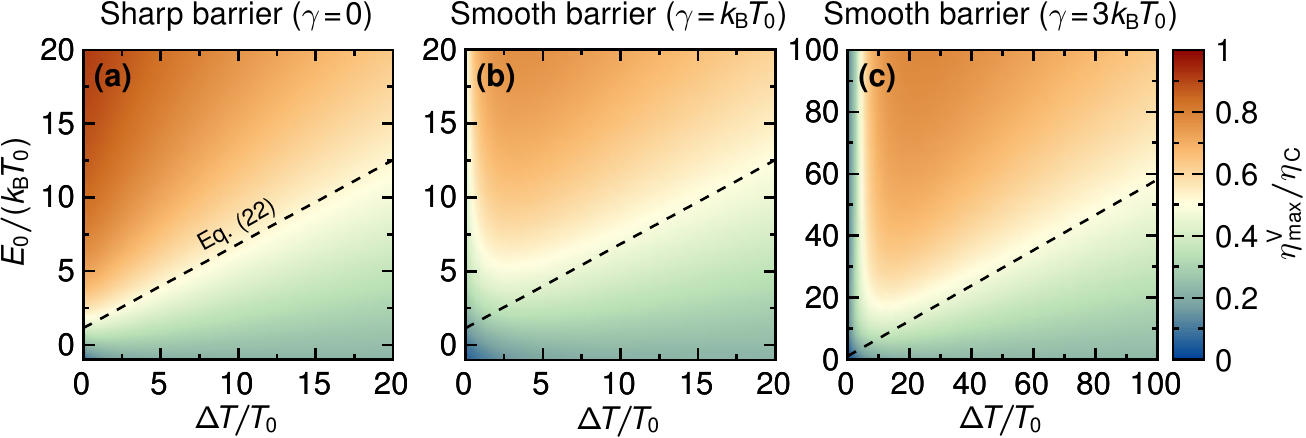}
\captionof{figure}{Density plots of the maximum efficiency~$\eta_\text{max}^\text{V}$ (that is, maximized over the voltage $V$) as a function of temperature gradient $\Delta T$ and step energy $E_0$, for three different values of barrier smoothness, $\gamma/k_\text{B}T_0=0,1,3$. Note that in panel (c) the axes ranges are strongly enlarged. 
\label{fig:powsharp}}
\end{center}
\end{figure}

We now focus our study on the maximum value of the efficiency that can be reached over the whole range of voltages, $\eta_\text{max}^\text{V}$, a function of $E_0/(k_\text BT_0),\Delta T/T_0, \gamma/(\Delta T/T_0)$. The results of this maximization procedure are shown in Fig.~\ref{fig:powsharp}, where panel~(a) corresponds to a sharp barrier ($\gamma=0$) while smooth barriers with $\gamma=k_\text{B}T_0$ and $\gamma=3k_\text{B}T_0$ are presented in panels (b) and (c), respectively.

Two important results can be immediately seen from these density plots of the efficiency~$\eta_\text{max}^\text{V}$ as a function of temperature gradient~$\Delta T/T_0$ and step energy~$E_0/(k_\text{B}T_0)$. First, we confirm the observations about the response to small temperature gradients made from Fig.~\ref{fig:eff}. While for a sharp barrier, efficiencies close to Carnot efficiency are reached in the linear response (close to the stopping voltage, as we know from Fig.~\ref{fig:eff}), for even only slightly smoothed barriers this is not the case anymore. For $\gamma/k_\text{B}T_0=3$, the maximum efficiency in the linear response is even suppressed down towards zero. This clearly shows that whenever the barrier step is not truly sharp, \textit{non-linear response} is required in order to get a thermoelectric response with large power output and with high efficiency. Second, panels~(b) and~(c) of Fig.~\ref{fig:powsharp} show that for temperature gradients much larger than the smoothness---or, in other words, with one of the reservoir temperatures being much larger than the smoothness---almost the same (large) efficiency as in the sharp-barrier case is found, as long as the step energy is sufficiently large. Note, however, that these large-efficiency regions are \textit{far} from those regions, which were previously identified as the ones of large power output, and are furthermore limited to regions with very large temperature gradients and step energies.

\subsection{Power-efficiency relations}

\begin{figure}[t!!!]
\centering
\includegraphics[scale=1]{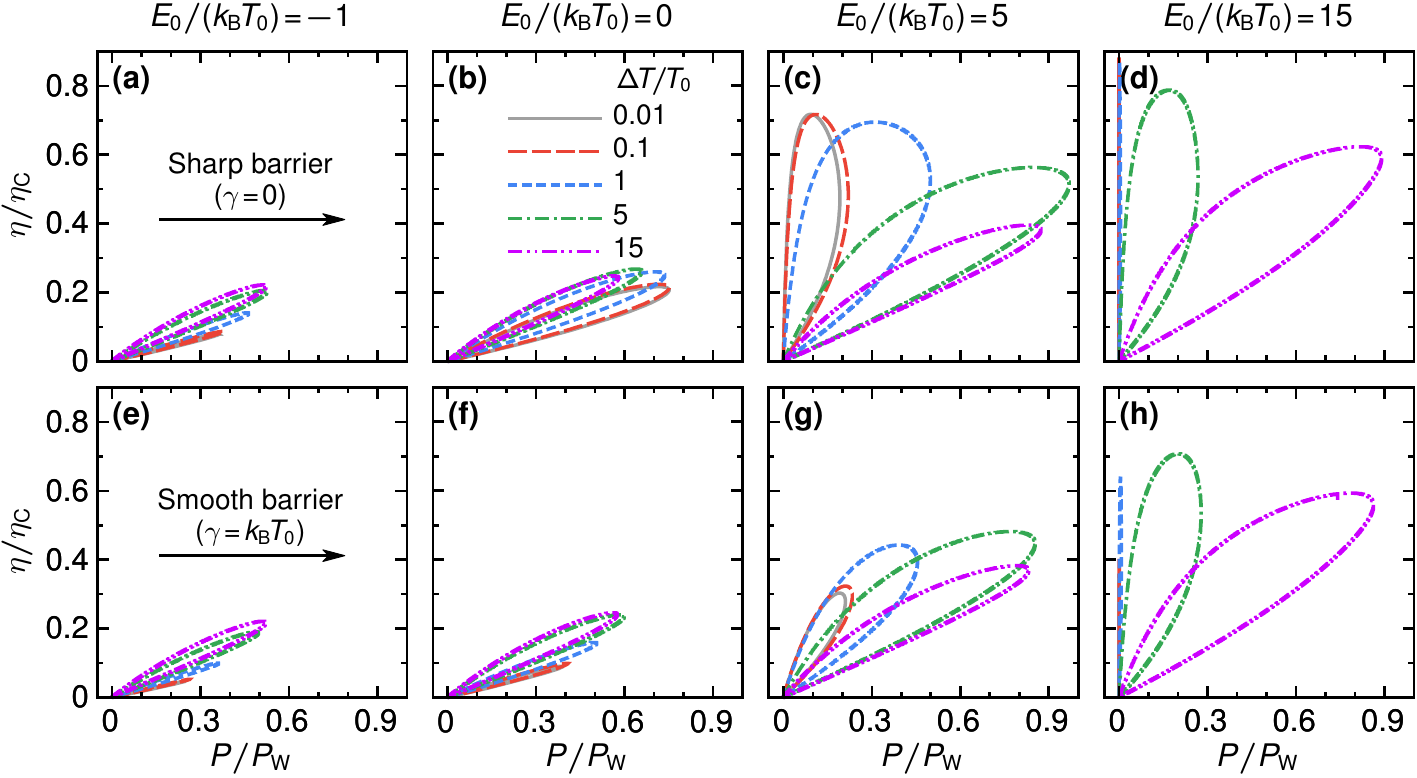}
\captionof{figure}{So-called lasso diagrams, showing the efficiency at every power output. The parameter that is changed along the lasso-line is the applied voltage $V$. We show results for a sharp barrier (top row) and for a barrier with smoothness $\gamma=k_\text{B}T_0$ (bottom row), for selected values of the step energy $E_0/(k_\text{B}T_0)$ (see different columns) and temperature gradients $\Delta T/T_0$ (see different lines), in analogy to Figs.~\ref{fig:pow-line} and~\ref{fig:eff}. 
\label{fig:lasso}
}
\end{figure}

The relation between power and efficiency for a sharp barrier, $\gamma=0$, was investigated in great detail in Refs. \cite{Whitney2014Apr,Whitney2015Mar,Whitney2013Aug,Pilgram2015Nov}. A convenient way to present the efficiency at a given power output, and vice versa, is in the form of lasso diagrams, as shown in Fig.~\ref{fig:lasso}.

At small step energies, $E_0/(k_\text{B}T_0)=-1,0$, the maximum power as well as the maximum efficiency are relatively small. However, maximum efficiency and maximum output power basically happen at the same parameter values. This is advantageous for operation of a thermoelectric device, where one typically has to decide whether to optimize the engine operation with respect to efficiency or power output.

This trend continues also for larger step energies, see panels (c) and (d) of Fig.~\ref{fig:lasso}, as long as the temperature gradient is larger or of the order of the step energy, $k_\text{B}\Delta T\gtrapprox E_0$ (meaning that  \mbox{$T_0+\Delta T>E_0/k_\text{B}>T_0$}).  In this case, the power output is close to its maximum value $P\approx P_\text{W}$, while the efficiency still takes values of up to the order of $\eta\approx0.6\,\eta_\text{C}$, in agreement with the bounds discussed in Refs. \cite{Whitney2014Apr,Whitney2015Mar,Pilgram2015Nov}. These results clearly show the promising opportunities of step-shaped energy-dependent transmissions, as they can possibly be realized in QPCs, for thermoelectric power production.

Note that the impressively large values for the efficiency at maximum output power do not, however, violate the Curzon-Ahlbohrn~\cite{Curzon1998Jun} bound, $\eta_\text{CA}$, which relates to the Carnot efficiency as
\begin{equation}
    \eta_\text{CA}=\frac{\eta_\text{C}}{1+\big(1+\Delta T/T_0\big)^{-1/2}}\ .
\end{equation}
This predicts a bound on the efficiency at maximum power of $\eta_\text{CA}=0.5\,\eta_\text{C}$ in linear response in $\Delta T$. That this bound is respected, can for example be verified by noting that the efficiency at maximum power of the grey solid line for~$\Delta T/T_0=0.01$ in panel (c) is only slightly above $0.4\eta_\text{C}$. Equally, one can check from the green dashed-dotted line in the same panel that the efficiency at maximum power does not exceed the bound for $\Delta T/T_0=5$ given by $\eta_\text{CA}=0.7\, \eta_\text{C}$.

For step energies that are large with respect to the temperature of \textit{both} reservoirs, $T_0,T_0+\Delta T<E_0/k_\text{B}$, the power output is reduced, the maximum efficiency, however, increases. In the limit of linear response in the temperature gradient, efficiencies close to Carnot efficiency are reached at the expense of close-to-zero power output.

\section{Power fluctuations and inverse Fano factor}\label{sec_fluc}
\begin{figure}[b!!!]
\centering
\includegraphics[scale=1]{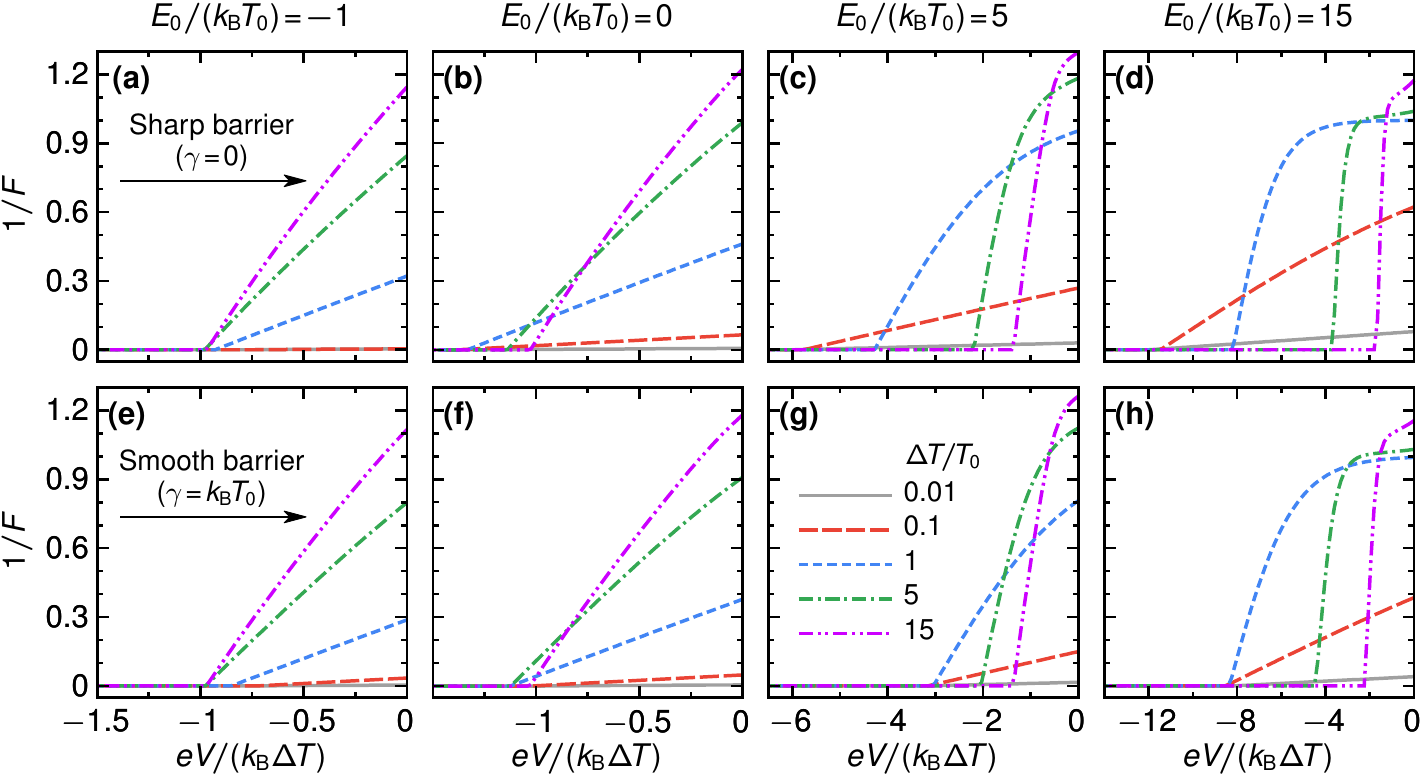}
\captionof{figure}{Inverse Fano factor as a function of voltage for sharp barrier (top row) and for smooth barrier (bottom row), for selected gradients $\Delta T$ (see different lines) and step energies $E_0$ (see different columns). Note that we set the inverse Fano factor to zero outside the parameter regime where power is produced. }
\label{fig:fanov}
\end{figure}
\begin{figure}[t!!!]
\centering
\includegraphics[scale=1]{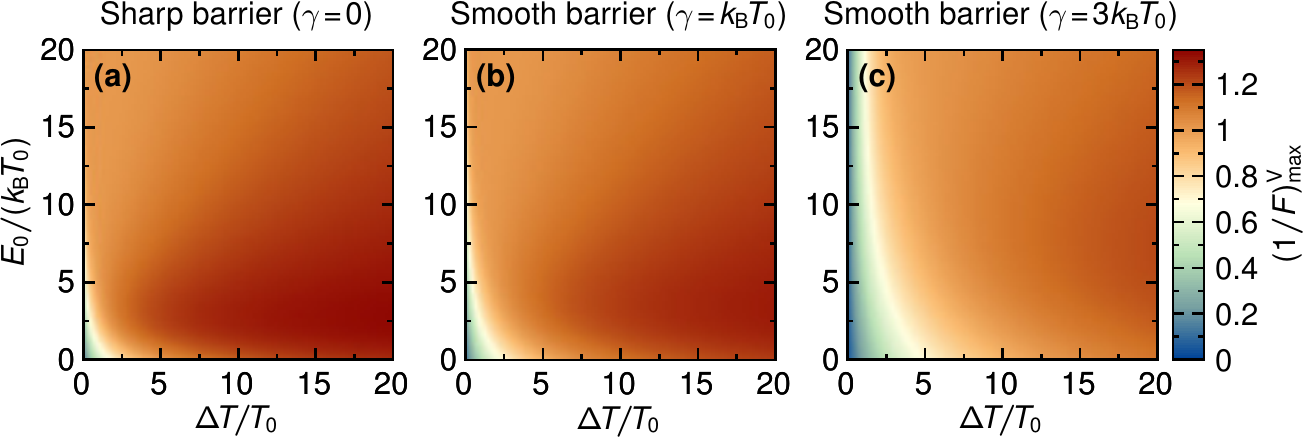}
\captionof{figure}{The inverse Fano factor  maximized over all those bias values $V$ leading to a non-negative output power, $\left(1/F\right)^\text{V}_\text{max}$, as a function of temperature gradient $\Delta T$ and step energy $E_0$, for three different values of barrier smoothness, $\gamma/k_\text{B}T_0=0,1,3$.
}
\label{fano}
\end{figure}

During recent years it has become clear that in addition to the power and the efficiency as performance indicators of a heat engine, the fluctuations of the power output, $S_P$, should also be considered. A reliable operation of the heat engine, i.e., where fluctuations are limited, is desirable. This is particularly relevant for nanoscale devices, where fluctuations are always a sizable effect. To analyze the effect of power fluctuations, we note that the relevant fluctuations in this QPC steady-state thermoelectric heat engine are the charge current fluctuations, since $S_P=V^2S_I$. Therefore, we shift the analysis of power fluctuations to the more straightforward analysis of the Fano factor, see Eq.~(\ref{eq_fano}). 
In~Fig.~\ref{fig:fanov}, we plot the inverse Fano factor $1/F$ as a function of voltage~$eV/(k_\text{B}\Delta T)$ for different barrier smoothness~$\gamma$, thermal gradients $\Delta T/T_0$, and step energies $E_0/(k_\text{B}T_0)$. Note that we set the inverse Fano factor to zero outside the parameter range where power is produced, in order to be able to use it as a performance quantifier. This performance quantifier $1/F$ is desired to be large, meaning that current fluctuations are small with respect to the average. For all parameters, we find that increasing the (negative) voltage decreases the inverse Fano factor $1/F$ (meaning that the Fano factor $F$ increases). This behavior is attributed to the decrease in charge current as the voltage is moved closer to the stopping voltage $V_\text s$. For small voltages, as well as small and negative step energies, increasing the thermal gradient generally increases the inverse Fano factor. These results can be understood from the linear response expression
\begin{equation}
    \label{eq:invfano}
    \frac{1}{F} = \left|\frac{eV}{k_{\rm B} T_0}+\frac{eL}{k_{\rm B}G}\cdot\frac{\Delta T}{T_0}\right|,
\end{equation}
where the absolute value can be omitted when focusing on the voltage window in which power is produced. This expression increases with $\Delta T$ and decreases as $V$ goes to more negative values. Increasing $\Delta T$ thus increases the current without an accompanied increase in fluctuations because $S_I=2k_{\rm B}T_0G$ is independent of the bias in the linear response. For large step energies $E_0$, the inverse Fano factor no longer increases monotonically in $\Delta T$ but a non-monotonic behavior is observed, indicating a more subtle interplay between the fluctuations and the mean value of the current. We note that for almost all parameter values, the inverse Fano factor is substantially smaller than one which can be attributed to the relatively large thermal noise in the present system.

In Fig. \ref{fano}, the inverse Fano factor maximized over the voltage, $\left(1/F\right)^\text{V}_\text{max}$,  is shown for the same parameters as used in Figs. \ref{fig:powmax2D} and \ref{fig:powsharp}. Note that the maximization only includes the voltage window where positive electrical power is produced. We find that, for all three values of smoothness, the maximum inverse Fano factor increases monotonically with increasing $\Delta T$, saturating at values a bit above unity. The Fano factor is thus slightly below unity, a signature of almost uncorrelated, close-to Poissonian, charge transfer (for Poissonian statistics, $F=1$). At small $\Delta T \lesssim T_0$, close to equilibrium, the noise is large even though the average electrical current is small. As noted above, this is purely due to thermal fluctuations, resulting in a small inverse Fano factor.  

\section{Thermodynamic uncertainty relation}\label{sec_TUR}

We now turn to the investigation of the TUR, cf.~Eq.~\eqref{TUR1}, which provides a combined performance quantifier accounting for power output, efficiency and power fluctuations.
We first consider the TUR-coefficient $Q_\text{TUR}$ in the linear-response regime
\begin{figure}[t]
\centering
\includegraphics[scale=1]{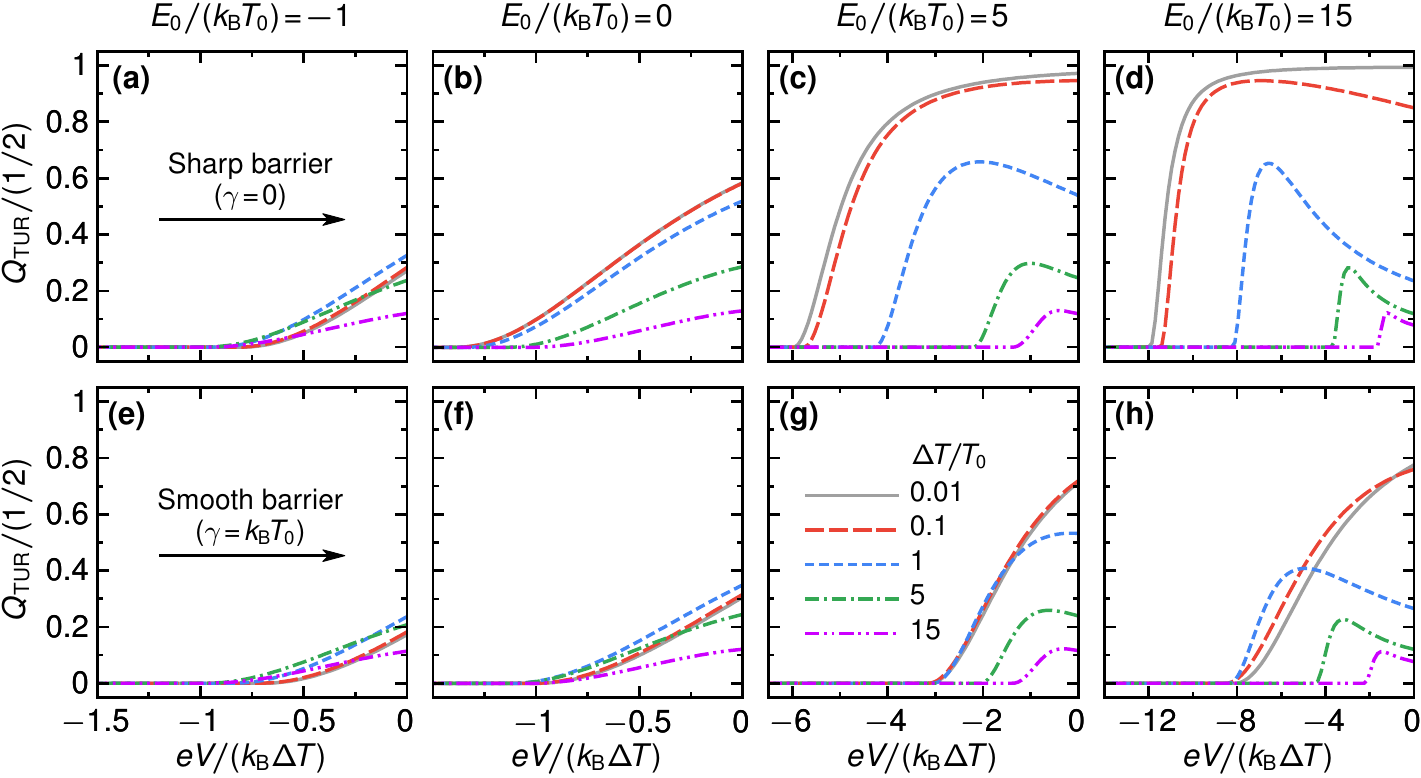}
\captionof{figure}{Coefficient $Q_{\rm TUR}$ as a function of voltage for sharp barrier (top row) and for smooth barrier (bottom row), for selected gradients $\Delta T$ (see different lines) and step energies $E_0$ (see different columns). We note that we set the $Q_{\rm TUR}$ to zero outside the parameter regime where power is produced. }
\label{fig:turv}
\end{figure}
%
\begin{equation}
Q_{\rm TUR}= \frac{(GV+L\Delta T)^2}{\Delta T(LT_0V+K \Delta T)+VT_0(GV+L\Delta T)}
\cdot
\frac{T_0}{2G}. 
 \label{TURlin}
\end{equation}
Maximizing this expression with respect to voltage we find $V_\text{max}\rightarrow \pm \infty$, resulting in $Q_{\rm TUR}=1/2$, and hence, the inequality becoming an equality. However, this voltage is outside the voltage regime for non-zero power production, \mbox{$V_\text{s}\leq V \leq 0$}. Thus, this is not of practical relevance for the engine performance. Adding the extra condition that $P\geq 0$ we instead find $V_\text{max}\rightarrow 0$. The corresponding value of $Q_\text{TUR}$ on the left hand side of Eq.~(\ref{TURlin}) then becomes $L^2T_0/(2GK)$. Expressing this in terms of the figure of merit $ZT$, given in Eq. (\ref{figofmerit}), we can write the bound on the operationally meaningful TUR-coefficient in the linear response regime as
\begin{equation}
 Q_{\rm TUR}\leq\frac{1}{2}
 \cdot
 \frac{ZT}{1+ZT}.
 \label{TURlin2}
\end{equation}
This shows that in the linear response, the parameters of the steady-state thermoelectric heat engine are actually subjected to a tighter bound than given by Eq.~(\ref{TURlin}). Note that this bound is saturated in the limit $V=0$, where the power production, the power fluctuations, as well as the efficiency all vanish. Also, only for $ZT \rightarrow \infty$, i.e., at the Carnot point, does the bound become $1/2$. As seen in Fig.~\ref{fig:turv}(d), this occurs for large step energies $E_0$.

\begin{figure}
\centering
\includegraphics[scale=1]{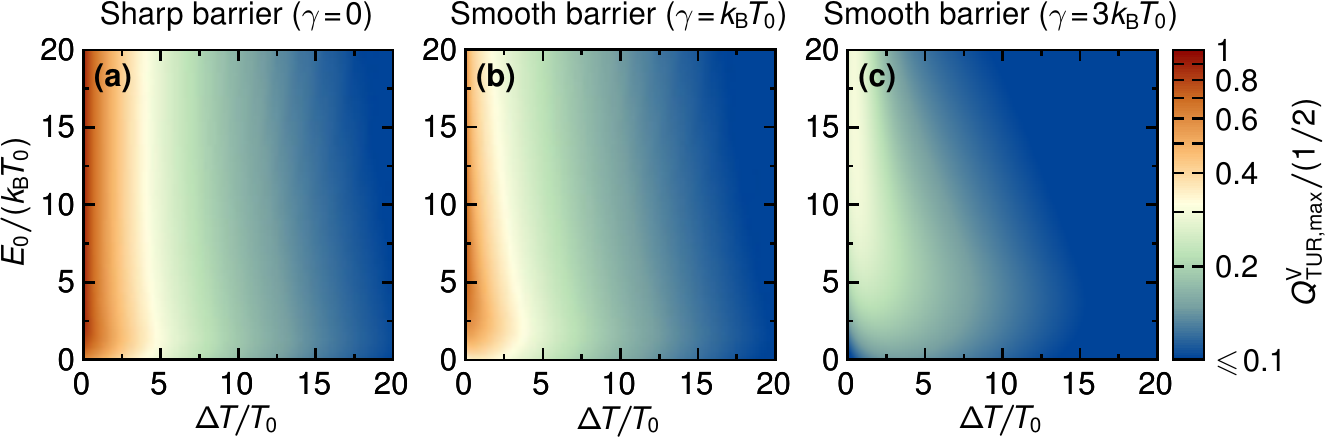}
\captionof{figure}{TUR-coefficient maximized over the bias $V$, $Q_{\rm TUR, max}^\text{V}$, as a function of temperature gradient $\Delta T$ and step energy $E_0$, for three different values of barrier smoothness, $\gamma/k_\text{B}T_0=0,1,3$.
}
\label{fig:TURs}
\end{figure}

The full TUR-coefficient beyond linear response is illustrated in Figs.~\ref{fig:turv} and \ref{fig:TURs}. We find that the inequality $Q_{\rm TUR}\leq1/2$ is always respected, even though this is not guaranteed by scattering theory~\cite{Agarwalla2018Oct}. Interestingly, we find the tighter bound in Eq.~\eqref{TURlin2} to be respected for most parameters, even though it is only proven to hold in the linear-response regime. Only for sufficiently low $E_0$ do we observe small violations of Eq.~\eqref{TURlin2} beyond the linear response regime [cf.~Fig.~\ref{fig:turv}~(a), (e), and (f)]. This is in agreement with the general notion that dissipation increases when moving away from the linear response \cite{Gingrich2016}. From Fig.~\ref{fig:turv}, we find that in the linear response, as well as for small and negative $E_0$, $Q_{\rm TUR}$ decreases monotonically as the (negative) voltage is increased. This reflects the behavior of the inverse Fano factor in Fig.~\ref{fig:fanov}. Importantly, for sharp step energies $E_0$, and beyond the linear response, $Q_{\rm TUR}$ is a non-monotonic function of the voltage and takes on its maximum at a point where power production is finite. This non-monotonicity is a consequence of the interplay between the monotonically decreasing inverse Fano factor and the strongly increasing efficiency and power [cf.~Figs.~\ref{fig:pow-line} and \ref{fig:eff}], as the voltage is changed to more negative values.

Figure \ref{fig:TURs} shows the TUR-coefficient maximized over voltage, $Q_{\rm TUR, max}^\text{V}$, as a function of the thermal gradient $\Delta T$ and the step energy $E_0$. As for the inverse Fano factor, the maximization only includes the voltage window where power is produced. For all values of the barrier smoothness, we find that $Q_{\rm TUR, max}^\text{V}$ generally decreases as a function of $\Delta T$, and a closer inspection reveals small non-monotonic features related to the small violations of Eq.~\eqref{TURlin2}. This is in contrast to the inverse Fano factor, which shows the opposite behavior, cf.~Fig.~\ref{fano}. The decrease of the fluctuations with $\Delta T$ is thus overcompensated by an increase in dissipation which results in the highest values for $Q_{\rm TUR, max}^\text{V}$ being reached in the linear response regime. This shows that $Q_{\rm TUR, max}^\text{V}$ is maximal in regimes, where the efficiency is large. In contrast, no features of the line of optimal power production close to $P_\text{W}$ can be identified in the panels of Fig~\ref{fig:TURs}. 

\section{Conclusions and Outlook}\label{sec_conclude}
In this paper, we have reviewed and extended the analysis of a QPC (or a QPC-like) device, with a transmission probability with a smoothed step-like energy-dependence, as a steady-state thermoelectric heat engine. The interest in a QPC for heat-to-power conversion derives from its optimal performance with respect to the output power, which goes along with rather large efficiencies.  We have analyzed the influence of the barrier smoothness on this behavior and found that strong non-linear-response conditions are required in order to recover a comparable performance.

In addition to the typically studied performance quantifiers---output power and efficiency---we have broadened the analysis by adding the \textit{power fluctuations} as an independent quantification of performance. The bound on the combination of these three quantities set by the recently identified thermodynamic uncertainty relation, suggests to investigate this as a combined performance quantifier.

We have shown that the bound of the thermodynamic uncertainty relation is further restricted if one adds the practical constraint of finite (positive!) output power. In the linear response, we quantify this restriction  by the figure of merit $ZT$. Interestingly, we have found that the combined performance quantifier has large values in those parameter regions in which the efficiency is maximized, while regions of maximal output power are not distinguished.

Whether this result is unique to the QPC as steady-state heat engine or can be generalized for other thermoelectric devices is a topic of further studies.

\authorcontributions{Numerical maximization of different quantities were mostly performed by S.K. and P.S., analytical estimates were mainly performed by N.D.,M.M., and P.S. All authors contributed to the discussion of results and the preparation of the manuscript.}

\funding{This research was funded by the Knut and Alice Wallenberg Foundation via an Academy Fellowship (N.D.,M.M.,J.S.), by the Swedish VR (S.K.,J.S., P.S.), and by the European Union's Horizon 2020 research and innovation programme under the Marie Sk{\l}odowska-Curie Grant Agreement No. 796700 (P.P.P.).}

\acknowledgments{We acknowledge fruitful discussions with A. Svilans and H. Linke.}

\conflictsofinterest{The authors declare no interest in conflicts.}


\end{document}